\documentclass{iau}
\usepackage{graphicx} 
\usepackage{natbib}

\def\apjl{ApJL}
\def\apj{ApJ}
\def\mnras{MNRAS}
\def\aap{A\&A}
\def\nat{Nature}

\def\aj{AJ}

\title[IAUS291.~~The PALFA Survey] 
{The PALFA Survey: Going to great depths to find radio pulsars} 

\author[P. Lazarus et al.]  
{P. Lazarus$^1$
 on behalf of the PALFA Collaboration$^2$}

\affiliation{$^1$Max-Planck-Institut f\"ur Radioastronomie, \\ Auf dem H\"ugel
69, Bonn, Germany, 53121 \\ email: {\tt plazarus@mpifr-bonn.mpg.de}
\\[\affilskip]
$^2$for full listing of collaboration members please see
http://www.naic.edu/$\sim$deneva/palfa/}

\pubyear{2012}
\volume{291}  
\jname{\mbox{Neutron Stars and Pulsars: Challenges and Opportunities after 80
years}} \editors{J. van Leeuwen, ed.} 
\begin{document}

\maketitle

\begin{abstract}
The on-going PALFA survey is searching the Galactic plane ($|b| < 5^{\circ}$,
$32^{\circ} < l < 77^{\circ}$ and $168^{\circ} < l < 214^{\circ}$) for radio
pulsars at 1.4 GHz using ALFA, the 7-beam receiver installed at the Arecibo
Observatory. By the end of August 2012, the PALFA survey has discovered 
100 pulsars, including 17 millisecond pulsars (P $<$ 30~ms).  Many of these
discoveries are among the pulsars with the largest DM/P ratios, proving that
the PALFA survey is capable of probing the Galactic plane for millisecond
pulsars to a much greater depth than any previous survey. This is due to the
survey's high sensitivity, relatively high observing frequency, and its high
time and frequency resolution. Recently the rate of discoveries has increased,
due to a new more sensitive spectrometer, two updated complementary search
pipelines, the development of online collaborative tools, and access to new
computing resources. Looking forward, focus has shifted to the application of
artificial intelligence systems to identify pulsar-like candidates, and the
development of an improved full-resolution pipeline incorporating more
sophisticated radio interference rejection. The new pipeline will be used in a
complete second analysis of data already taken, and will be applied to future
survey observations. An overview of recent developments, and highlights of
exciting discoveries will be presented.

\end{abstract}


\firstsection 
\section{Introduction}
The PALFA survey is an on-going, large-scale, survey for radio pulsars tuned to
find millisecond pulsars (MSPs). PALFA is similar to, but more sensitive than
other on-going projects (e.g. HTRU-N/S, GBNCC, SPAN512, described by Ng et al.,
Keith et al., Lynch et al., and Desvignes et al., respectively, in these
proceedings). MSPs are particularly useful for searches for the direct
detection of nano-Hz gravitational waves using Pulsar Timing Arrays
\citep[e.g.][]{vlj+11}, strong-field tests of relativistic theories of gravity
\citep[e.g.][]{ksm+06}, studying the equation of state of ultra-dense matter
\citep[e.g.][]{dpr+10}.  Another main goal of the survey is to expand the known
population of radio pulsars, to gain a better understanding of their overall
numbers, the distribution of their properties, determine birth-rates, and
estimate birth properties. The survey is sensitive to transient pulsars, such
as nullers, rotating radio transients (RRATs), intermittent pulsars, and radio
magnetars.  Discoveries may also be young pulsars associated with high-energy
emission, supernova remnants, or pulsar wind nebulae. PALFA has the potential
for discovering even more exotic systems such as pulsar-black hole binaries,
which could have profound ramifications for the understanding of gravitation
\citep[e.g.][]{wk99}.

The PALFA Survey makes use of the 7-beam L-band (1.4 GHz /
21-cm) receiver installed on the Arecibo Observatory's William E. Gordon
Telescope. Observing has focused on the plane of the Galaxy ($|b| <
5^{\circ}$), in the two regions visible from Arecibo, due to its transit
design, the ``inner Galaxy'' ($34^{\circ} < l < 77^{\circ})$ and the
``anti-centre'' ($168^{\circ} < l < 214^{\circ}$) regions.

The Wideband Arecibo Pulsar Processor (WAPP) spectrometers used by the PALFA
survey since its beginning in 2004 were replaced with the more sensitive Mock
spectrometer system in 2009. The Mock set-up allows for ALFA's full 320 MHz
bandwidth to be recorded, more than tripling the bandwidth available compared
to the older WAPP set-up. In 2009, before completely phasing out the WAPP
spectrometers for use in the survey, both backends were run in parallel to
ensure the quality of the data from the new system. The ALFA receiver's system
temperature is $\sim$24 K, and the gain is $\sim$10.4 K/Jy for the central beam
and $\sim$ 8.2 K/Jy for the outer beams \citep{cfl+06}. The survey's observing
parameters, summarized in Table \ref{tab:params}, are characterized by high
time and frequency resolution. Integration times are 4.5 min for
the inner Galactic region, and 2.25 min for the anti-centre region.

\begin{table}[t]
\footnotesize
\centering
\caption{PALFA Survey observing system parameters.}
\label{tab:params}
\begin{tabular}{rcc}
    \hline
    & WAPP Spectrometers & Mock Spectrometers \\
    \hline
    Centre Frequency & 1420/1440 MHz$^{a}$ & 1375.432 MHz \\
    Sample time & 64 $\mu$s & 65.476 $\mu$s \\
    Bandwidth & 100 MHz & 322.617 MHz \\
    Number of Channels & 256 & 960 \\
    \hline
\end{tabular}
\footnotesize
\\$^{a}$ The centre frequency of the WAPPs was increased in 2005 Nov. to reduce interference.
\end{table}

The high resolution employed by the PALFA survey allows for the dispersive
smearing caused by free electrons in the interstellar medium along the
line-of-sight to be almost completely removed for pulsars at any depth into the
Galactic plane using incoherent dedispersion techniques. Scattering due to
multi-path propagation still limits the depth to which the fastest spinning
pulsars can be detected within the Arecibo-visible sky (see Figure
\ref{fig:broad+timeline}, left).

The large collecting area and hence high gain of the Arecibo Observatory make
the PALFA survey the most sensitive large-scale survey of the Galactic plane to
date. The high sensitivity of the survey enables integration times far shorter
than would be possible when using smaller telescopes, and reduces the total
data volume of the survey. It is important to note that shorter observations
are more computationally efficient to search, especially for pulsars in binary
systems.

\section{Processing \label{sec:processing}}
Processing of PALFA survey data is currently done with two complementary
pipelines. One is based on the PRESTO suite of pulsar search code
\footnote{http://www.cv.nrao.edu/$\sim$sransom/presto/}. The other pipeline
makes use of the BOINC-based\footnote{http://boinc.berkeley.edu/index.php}
Einstein@Home distributed global volunteer computing
platform\footnote{http://einstein.phys.uwm.edu/}.

Before data are distributed to processing sites the data are converted to
4-bits-per-sample PSRFITS format, and archived at Cornell University's Center
for Advanced Computing. All data are tracked by a database at Cornell. Data are
automatically downloaded from the archive by processing sites.

\subsection{PRESTO}
The PRESTO pipeline first searches for, and removes, bright un-dispersed
narrow-band impulsive, and periodic, radio frequency interference (RFI) signals
in the data.

The pipeline then dedisperses the data with an assumed dispersion measure (DM)
value, removing the corresponding frequency-dependent delay. Since the DM of
unknown pulsars are not known \textit{a priori}, 4188 (1140) dedispersed time series,
each with a different trial DM value ranging from 0 to $\sim$1000 pc~cm$^{-3}$,
are generated when searching Mock (WAPP) observations. Each resulting time
series is then searched in the time domain for bright individual pulses, and
in the Fourier domain for periodic signals.

Each dedispersed time series is searched for single pulses. To increase
sensitivity to pulses of width larger than the observation's sampling time a
series of boxcar matched filters of different widths are convolved with the
data. A summary plot of all significant candidate pulses identified at all
trial DMs is produced, and inspected.

In order to increase sensitivity to pulsars in binary systems a series of trial
Fourier-domain templates, each corresponding to a constant acceleration, are
used when searching the Fourier Transform of the dedispersed time series
\citep[see][for more details]{rem02}. Given the length of PALFA observations,
and the number and range of trial acceleration values, the PRESTO pipeline
maintains sensitivity to pulsars in binary orbits with periods as small as
$\sim$1~hour.

The PRESTO pipeline then identifies the most likely pulsar candidates from the
list of periodicity signals identified. For each periodicity candidates the raw
data are folded using its period and DM, and a diagnostic plot is produced. Also,
a series of scoring heuristics are computed for each folded candidates. These
are later used as input to Artificial Intelligence (AI) systems trained to identify
pulsar-like candidates, and ignore RFI-like and noise-like signals that were
folded by the pipeline. All periodic signals identified are summarized in
overview diagnostic plots.

The PRESTO pipeline is run on computing resources at the University of British
Columbia, and on the Guillimin
supercomputer\footnote{http://www.clumeq.ca/index.php/en/about/computers/guillimin}
managed by McGill University. Combined, 3000-6000 observations can be reduced
per month, depending on contention for resources on Guillimin.

\subsection{Einstein@Home}
Prior to distributing data to volunteers' computers, the Einstein@Home-based
pipeline identifies, and masks narrow-band impulsive RFI in the time-domain.
The pipeline also identifies periodic RFI, which is replaced by random noise.
The pipeline then generates 3808 (628) dedispersed time series for Mock (WAPP)
data, each with a different trial DM, up to $\sim$1000 pc~cm$^{-3}$. These
dedispersed time series that are transferred to volunteers' computers to be
analyzed.

Sensitivity to binary pulsars is improved by using 6662 circular orbital
templates to demodulate the time series in the time domain. This technique is
capable of detecting isolated pulsars \citep[e.g.][]{kac+10}, and still
maintains sensitivity to pulsars in binaries with orbital periods as short as
$\sim$11 min. The Fourier transform of each demodulated time series is searched
for periodic signals. The top 100 signals are returned to the Einstein@Home
servers where they are collated, diagnostics plots are created, and heuristic
pulsar-identification algorithms are run.

No single pulse search is performed in the Einstein@Home analysis.

The Einstein@Home pipeline can make use of volunteers' GPUs, if permitted.
Processing on GPUs accounts for $\sim$75\% of the pipeline's processing power.
The performance of the pipeline, and more importantly the GPU code, is
constantly being improved, reducing the time required for analysis with each
iteration. Currently, the Einstein@Home pipeline searches $\sim$3000
observations per month.


\subsection{Candidates}
The PALFA collaboration uses a centralized database to store metadata and
diagnostics for each observation, and details for candidates identified by
processing. The database holds approximately six million candidates.

Candidates are primarily inspected using an online viewer available on the
PALFA Survey's portal on the CyberSKA website.\footnote{www.cyberska.org/palfa.
Access is restricted to members of the PALFA Consortium.} The viewer allows
users to select candidates based on information in the database, including
scoring heuristics, AI recommendations, and the usual observation/candidate
information. The user interface provides one-click access diagnostic
information and plots about the candidate being viewed, and the observation in
which it was found. The candidate viewer also provides direct access to another
web-application designed to track the most promising candidates, and their
follow-up. Both applications, and others not mentioned here, are hosted at
McGill University.

\section{Results}
\begin{figure}[t]
    \centerline{
    \includegraphics[width=2.5in]{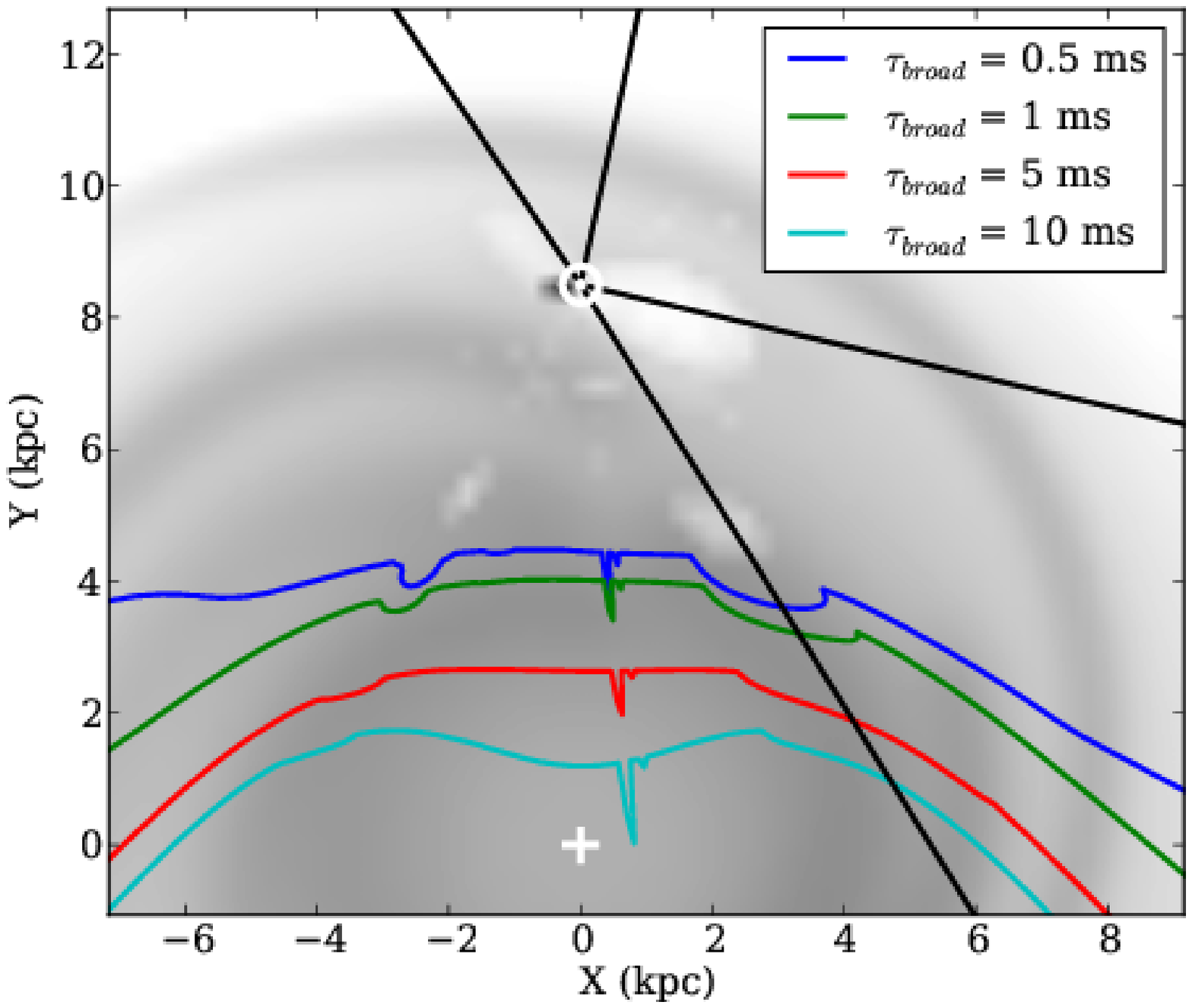}
    ~
    \includegraphics[width=2.75in]{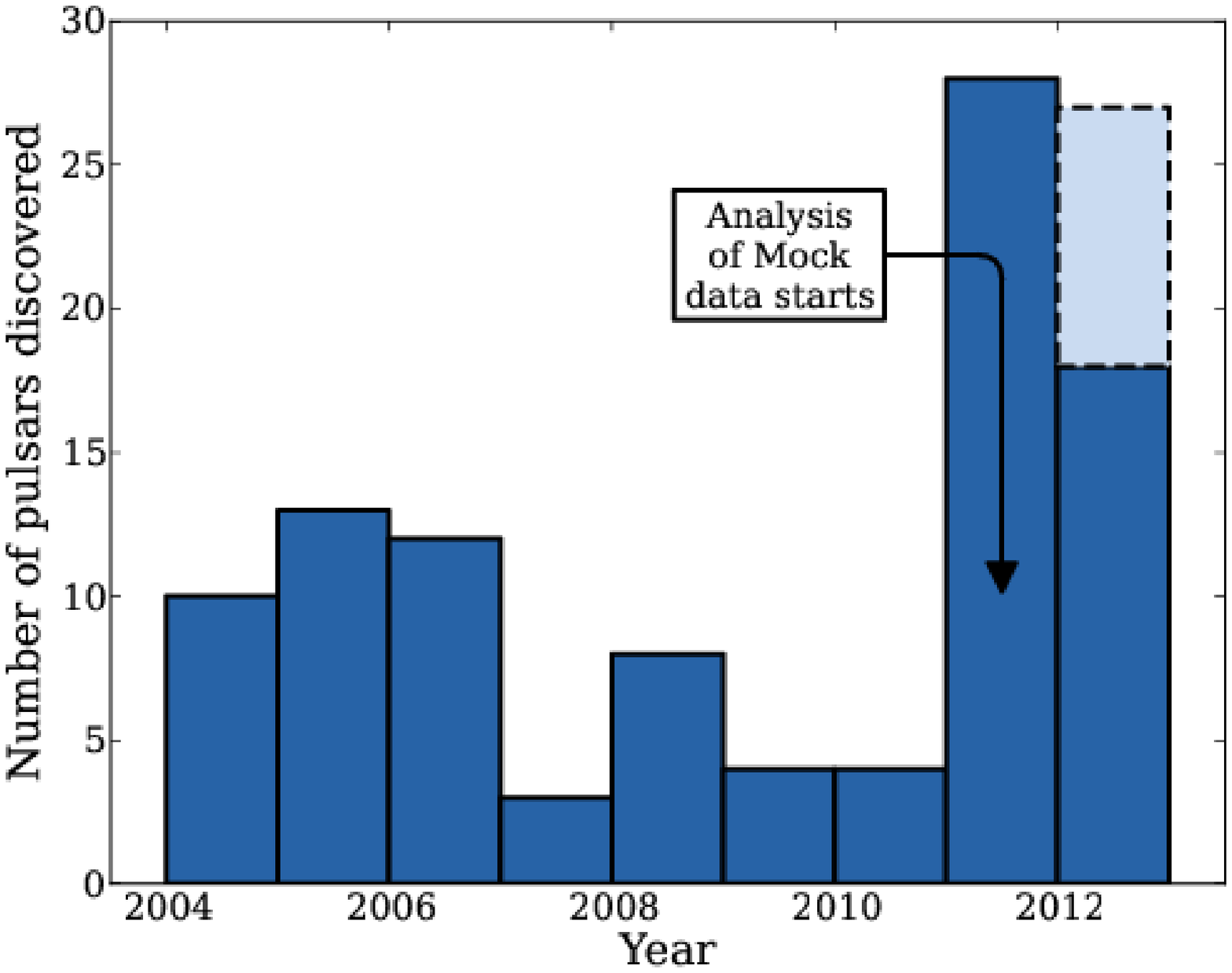}
}
\caption{{\it Left}.  A top-down view of the Galactic plane showing Galactic
MSP overlaid with the depths corresponding to various pulse broadening times
as predicted by the NE2001, for the PALFA survey's observing parameters. \quad
{\it Right}. Number of PALFA discoveries per year since the start of the
survey. The projected number of discoveries for 2012 is shown with a dashed
extension. \label{fig:broad+timeline}}
\end{figure}

By the end of August 2012, the PALFA survey has discovered 100 pulsars. The
discovery rate increase significantly when the analysis of the Mock data began
(see Figure \ref{fig:broad+timeline}, right). This is partly due to the
increase of computing resources available. However, the number of discoveries
per sq. deg. is higher for Mock data, owing to their higher sensitivity. The WAPP
spectrometers have been used to discover 56 pulsars in 164 sq. deg. of PALFA
data (1 pulsar per 2.9 sq. deg.) whereas 44 pulsars have been discovered in
only 70 sq. deg. of Mock data (1 pulsar per 1.6 sq. deg.).

\subsection{Highlights}
The PALFA survey has made several exciting discoveries since its start,
including:

\begin{description}
\item[PSR~J1906+0746] \quad (\textit{P=144~ms; DM=212~pc~cm$^{-3}$}) 
This binary pulsar has a characteristic age of
$\tau_{c}=\frac{P}{2\dot{P}}$=112~kyr, making it the youngest known binary
pulsar \citep{lsf+06}. Also, it is the second most relativistic binary pulsar
system known. \citep[See also,][]{kas12}.

\item[PSR~J1903+0327] \quad (\textit{P=2.15~ms; DM=298~pc~cm$^{-3}$}) 
This Galactic MSP is in a binary orbit whose eccentricity of $e=0.44$
challenges the standard MSP formation picture \citep{crl+08}.
Pulsar timing has measured the Shapiro delay, and the relativistic advance of
periastron, allowing the determination of the pulsar's mass
($M_p=1.67(2)M_\odot$) and the companion's mass ($M_c=1.029(8)M_\odot$). Near-IR
spectroscopic observations have identified the companion as a main-sequence
star. Simulations suggest the current binary was formed from a triple-star
system \citep{fbw+11,pvvn11}.

\item[PSR~J1856+0245] \quad (\textit{P=80.9~ms; DM=650~pc~cm$^{-3}$}) 
This young ($\tau_{c}=$ 27~kyr), highly energetic ($\dot{E}=4\pi^{2}I\dot{P}P^{-3}=4.6 \times
10^{36}$~erg~s$^{-1}$) pulsar is coincident with the TeV $\gamma$-ray source
HESS~1857+026, as well as a faint X-ray source AX~J185651+0245 \citep{hng+08,rgv+12}.

\item[PSR~J1949+3106] \quad (\textit{P=13.1~ms; DM=164~pc~cm$^{-3}$}) 
This binary MSP is in a circular ($e=4 \times 10^{-5}$), whose inclination
($i=80^{\circ}$) makes the detection of Shapiro delay possible in the timing
data \citep{dfc+12}. The measurement of Shapiro delay provides mass
measurements of the pulsar ($M_p=1.47^{+0.43}_{-0.31}~M_{\odot}$) and of the
companion ($M_c=0.85^{+0.14}_{-0.11}~M_{\odot}$). Detection of relativistic
advance of periastron, which could be possible within a few years, will greatly
reduce the uncertainties on the masses. Also, the absence of timing noise, the
decent timing precision ($RMS_{TOA} = 3.96\mu$s), and the as-of-yet unresolved
pulse profile, coupled with Arecibo's new PUPPI backend could make this pulsar
useful for Pulsar Timing Array experiments.
\end{description}

\subsubsection{Global distributed volunteer computing discoveries}
The Einstein@Home platform's first discovery was found in 2010 in a WAPP PALFA
data \citep[J2007+2722,][]{kac+10}. Since then, the Einstein@Home
pipeline has discovered 22 additional pulsars, including four MSPs, and three
binaries.

\subsubsection{Distant MSPs}
\begin{figure}[t]
    \centerline{
    \includegraphics[width=2.75in]{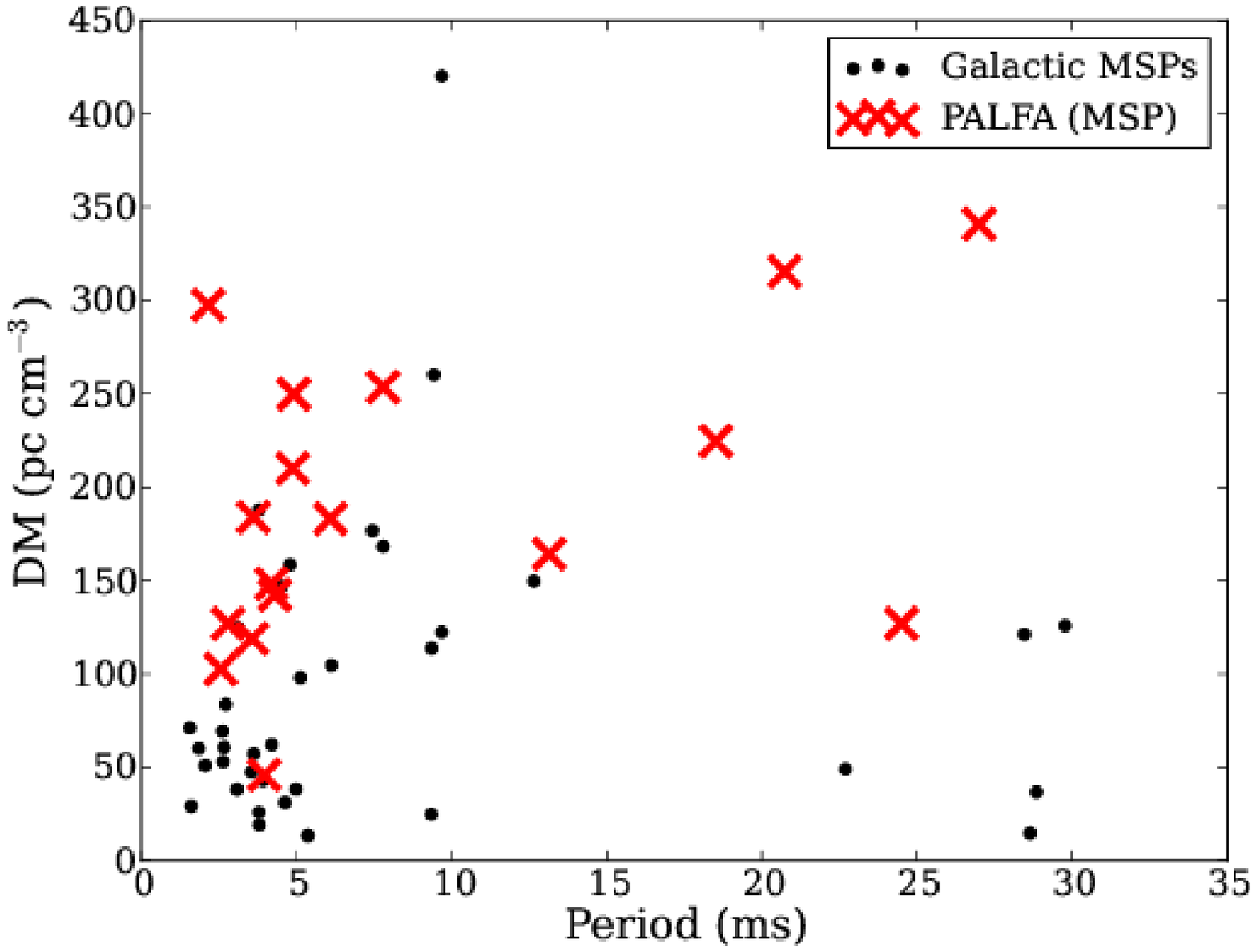} 
    ~
    \includegraphics[width=2.45in]{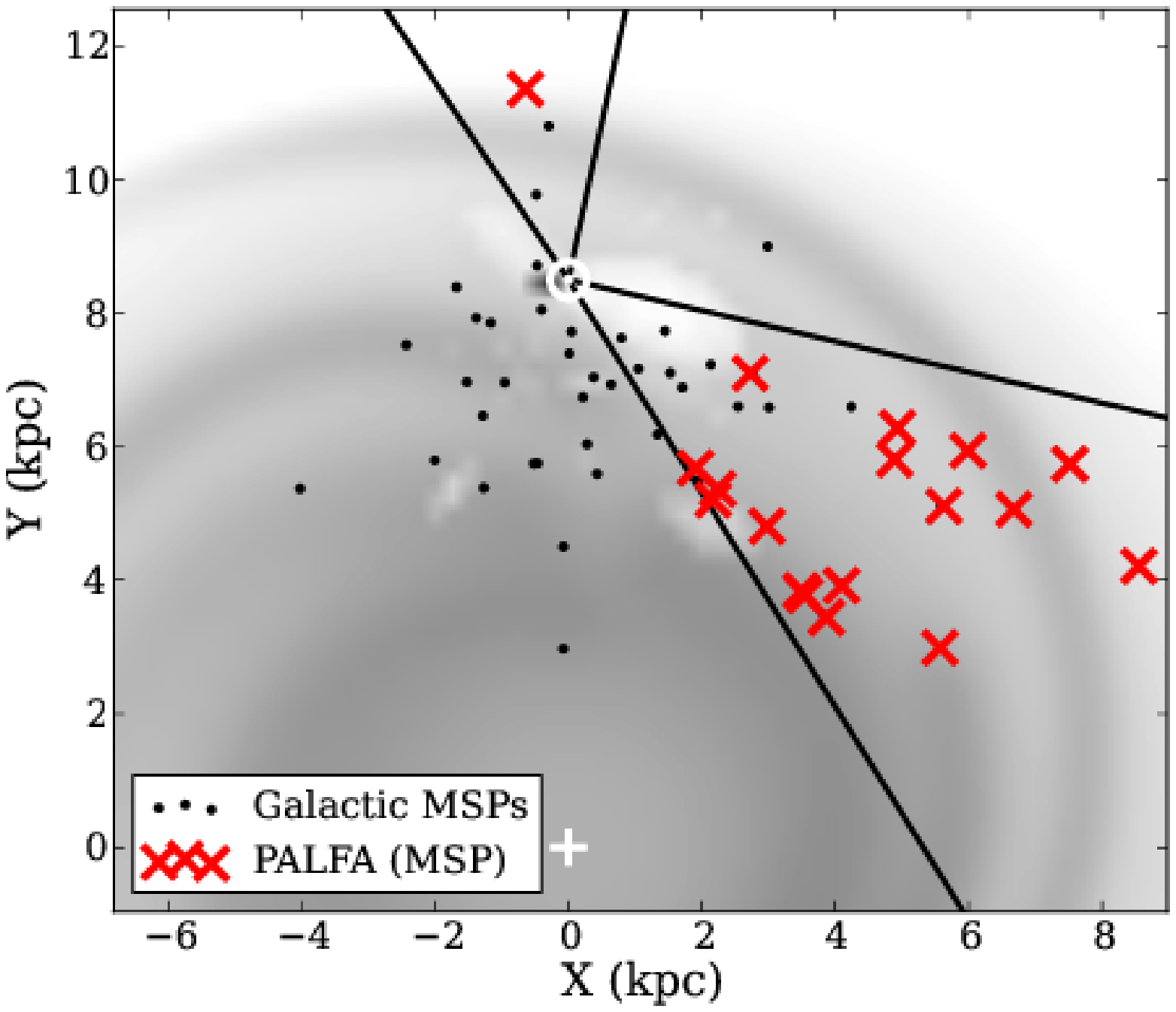}
}
\caption{{\it Left}. DM vs. Period for Galactic MSPs. Notice that red Xs,
PALFA-discovered MSPs tend towards the top of the plot. These pulsars are
generally thought to be harder to find. \quad {\it Right}. A top-down view of
the Galactic plane showing Galactic MSP. The PALFA-discovered MSPs (red Xs) are
generally located deeper into the plane of the Galaxy. Distances are inferred
from the DM using the NE2001 model. \label{fig:dmp+topdown}}
\end{figure}

The PALFA survey has discovered 17 MSPs (defined here as P $<$ 30~ms). These
pulsars are amongst the highest DM MSPs in the Galactic plane. The ratio DM/P
can be used as a rough measure of the difficulty of finding a pulsar, since the
smearing in time caused by DM is more detrimental to shorter period pulsars.
The MSPs found in the PALFA survey make up a large proportion of the highest
DM/P pulsars known (see Figure \ref{fig:dmp+topdown}, left).

Also, by assuming the NE2001 model for the Galactic distribution of
free-electron density \citep{cl02}, it is possible to infer distances given the
sky-position and DM of a pulsar. Figure \ref{fig:dmp+topdown} (right) shows the
inferred positions of PALFA-discovered MSPs projected onto the plane of the
Galaxy, compared to all other Galactic plane MSPs. The median distance to PALFA
MSPs is 5.8~kpc, double that of other known Galactic MSPs ($D_{median} =
2.9$~kpc), highlighting PALFA's ability to find distant MSPs.

\section{Future Outlook}
The PALFA survey has observed only 10\% of its nominal survey region with the
Mock spectrometers, which are proving to be superior to the WAPPs due to their
tripled bandwidth. The Mock spectrometers will be used as the survey continues,
and will eventually be used to re-observe regions covered with the WAPPs.

In addition, an updated version of the PRESTO pipeline, including enhanced RFI
mitigation, improved candidate optimization, additional diagnostic information,
and a few bug fixes, is being beta-tested on the Guillimin supercomputer. The
updated pipeline is already finding pulsars missed by the first analysis.
Improved single pulse analyses based on the work of \cite{sccs12} are also
being applied to the PALFA data taken thus far, and will be integrated into the
survey's standard data reduction, as will a modified version of the algorithm
developed by Karako-Argaman et al. (in these proceedings).

With the 100 pulsars discovered to date, the improved data quality, additional
computing resources available, and enhancement to the data reduction, the
PALFA survey is well on its way to discovering hundreds of pulsars, including
expanding the population of faint, distant MSPs it has already been successful
at finding.

\end{document}